\documentclass[oldversion]{aa}

\usepackage{times}
\usepackage{graphicx}
\usepackage[switch]{lineno}
\usepackage{xspace}
\usepackage{epsfig}
\usepackage{textcomp}
\usepackage{amsmath}
\usepackage{natbib}
\usepackage{tabularx}
\usepackage{rotating}
\usepackage{dcolumn}
\usepackage{epstopdf}

\bibpunct{(}{)}{;}{a}{}{,}

\newcommand{\isooph}{ISO-Oph\,85\xspace}

\def\gsim{\;\lower4pt\hbox{${\buildrel\displaystyle >\over\sim}$}\,}
\def\lsim{\;\lower4pt\hbox{${\buildrel\displaystyle <\over\sim}$}\,}
\begin{document}



\title{Results from DROXO}  

\subtitle{IV. EXTraS discovery of an X-ray flare from the Class\,I  protostar candidate \isooph}

\author{D. Pizzocaro\inst{\ref{inst1}, \ref{inst4}}, B. Stelzer\inst{\ref{inst2}}, R. Paladini\inst{\ref{inst15}}, A. Tiengo\inst{\ref{inst1}, \ref{inst6}, \ref{inst5}}, G. Lisini\inst{\ref{inst5}}, 
G. Novara\inst{\ref{inst6}}, G. Vianello\inst{\ref{inst8}}, A. Belfiore\inst{\ref{inst1}}, M. Marelli\inst{\ref{inst1}}, D. Salvetti\inst{\ref{inst1}}, 
I. Pillitteri\inst{\ref{inst2},\ref{inst10}}, S. Sciortino\inst{\ref{inst2},\ref{inst9}}, D. D'Agostino\inst{\ref{inst11}}, F. Haberl\inst{\ref{inst14}}, 
M. Watson\inst{\ref{inst12}}, J. Wilms\inst{\ref{inst13}}, R. Salvaterra\inst{\ref{inst1}}, A. De Luca\inst{\ref{inst1},\ref{inst6}}}

\offprints{}

\institute{
 INAF-Istituto di Astrofisica Spaziale e Fisica Cosmica Milano, via E. Bassini 15, 20133 Milano, Italy\label{inst1}  \\ \email{D. Pizzocaro, pizzocar@lambrate.inaf.it} \and
 Università degli Studi dell’Insubria, Via Ravasi 2, 21100 Varese, Italy\label{inst4} \and
  INAF - Osservatorio Astronomico di Palermo, Piazza del Parlamento 1, 90134 Palermo, Italy\label{inst2} \and
  Infrared Processing and Analysis Center, California Institute of Technology, Pasadena, CA 91125, USA \label{inst15}\and
  IUSS – Istituto Universitario di Studi Superiori, piazza della Vittoria 15, 27100 Pavia, Italy\label{inst6}\and
  INFN – Istituto Nazionale di Fisica Nucleare, Sezione di Pavia, via A. Bassi 6, 27100 Pavia, Italy\label{inst5}  \and
  W. W. Hansen Experimental Physics Laboratory, Kavli Institute for Particle Astrophysics and Cosmology, Department of Physics and SLAC National Accelerator Laboratory, Stanford University, Stanford, CA 94305, USA \label{inst8} \and
  Smithsonian Astrophysical Observatory (SAO) - Harvard Center for Astrophysics, Cambridge MA, USA\label{inst10} \and
  DSFA - Dipartimento di Scienze Fisiche e Astronomiche, Università degli Studi di Palermo, Piazza del Parlamento 1, 90134 Palermo, Italy\label{inst9} \and
  IMATI, Istituto di Matematica Applicata e Tecnologie Informatiche ``Enrico Magenes'', Via dei Marini 6, 16149 Genova, Italy\label{inst11} \and
  Max-Planck-Institut f\"{u}r extraterrestrische Physik, Giessenbachstrasse, 85748 Garching, Germany\label{inst14}\and
  Department of Physics and Astronomy, University of Leicester, Leicester LEI 7RH, UK\label{inst12} \and
  Dr. Karl-Remeis Sternwarte and ECAP, Universit\"{a}t Erlangen-N\"{u}rnberg, Sternwartstr. 7, 96049 Bamberg, Germany\label{inst13}\\
}

\titlerunning{\isooph}
\authorrunning{D. Pizzocaro et al.}

\date{Received $<$20-05-2015$>$ / Accepted $<$16-10-2015$>$}

\abstract{

\noindent X-ray emission from Young Stellar Objects (YSOs) is a key ingredient in understanding 
star formation. For the early, protostellar (Class\,I) phase, a very limited (and controversial) quantity of X-ray results is available to date. 

\noindent Within the EXTraS (\textit{Exploring the X-ray Transient and variable Sky}) project, 
we have discovered transient X-ray emission 
from a source whose counterpart is \isooph, a strongly embedded YSO in the $\rho$ Ophiuchi 
star-forming region.

\noindent We extract an X-ray light curve for the flaring state, 
and determine the spectral parameters for the flare
from {\em XMM-Newton}/EPIC data 
with a method based upon quantile analysis. 
We combine photometry from infrared to millimeter 
wavelengths from the literature with mid-IR {\em Spitzer} and  unpublished submm {\em Herschel} photometry 
that we analysed for this work, and we describe the resulting SED with a set of 
precomputed models.

\noindent The X-ray flare of \isooph lasted $\sim2500\,{\rm s}$ and  is consistent with 
a highly-absorbed one-component thermal model ($N_H=1.0_{-0.5}^{+1.2}\cdot10^{23}\,cm^{-2}$
and $kT=1.15_{-0.65}^{+2.35}\,{\rm keV}$). The X-ray luminosity during the flare is 
$\log L_X\,{\rm [erg/s]}=31.1^{+2.0}_{-1.2}$; during quiescence we set an upper limit of $\log L_X\,{\rm [erg/s]} < 29.5$.
We do not detect other flares from this source.
The submillimeter fluxes suggest that the object is a Class\,I protostar. We caution, however, that the offset between the {\em Herschel} and optical/infrared 
position is larger than that for other YSOs in the region, leaving some doubt on this association. 

\noindent To the best of our knowledge, this is the first X-ray flare from a YSO that has been recognised as a candidate 
Class\,I protostar via the analysis of its complete SED, including the submm bands
that are crucial for detecting the protostellar envelope. 
This work shows how the analysis of the whole SED is fundamental to the classification of 
YSOs, and how the X-ray source detection techniques we have developed can open a new era in 
time-resolved analysis of the X-ray emission from stars.}


\maketitle

\section{Introduction}\label{sect:intro}

X-ray emission that largely arises from a stellar dynamo is a prime characteristic and well-studied 
phenomenon of pre-main sequence stars,  both with and without accretion disks (Class\,II and Class\,III Young Stellar Objects (YSOs)\footnote{We adopted an infrared (IR) classification scheme
for young stellar objects devised by \cite{1987IAUS..115....1L} on the basis of the shape of their 
spectral energy distribution (SED).}). No standard dynamos are  
predicted for the earliest protostellar evolutionary stages of IR Class\,0 and I. 
Yet X-ray studies of young star clusters have repeatedly come up with the detection of a small 
fraction of Class\,I objects \citep[e.g.,][]{Prisinzano08.0, Guenther12.0} and a few 
reports of X-ray detections of borderline Class\,0/I sources, i.e., strongly embedded objects
without counterpart in the near-IR,
are also present in the literature \citep[e.g.][]{Hamaguchi05.2, Getman07.1}. 
These days, the most stringent upper limits on the X-ray emission from bona fide Class\,0
YSOs is $\log L_X <29.6$, derived by \cite{2007A&A...463..275G}
thanks to the analysis of a deep Chandra observation of the Serpens star-forming region. 
The lower number of X-ray detections  of Class\,0 and Class\,I 
protostars, with respect to Class\,II and Class\,III pre-main sequence stars, may be related to the strong 
extinction from the envelopes of the former ones, preventing the detection of the  soft ($\sim 1\,{\rm keV}$) X-rays 
that dominate the emission from Class II and Class III objects.
In fact, high gas column densities ($N_{\rm H} > 10^{22}\,{\rm cm^{-2}}$) have been
found for the small number of protostars with sufficient X-ray photons for spectral 
analysis \citep{ Pillitteri10.1, 2003ApJ...582..398F, 2007A&A...468..353G}. 
So far, low number statistics  have  hindered a conclusion on the
presence of intrinsic differences in the plasma temperatures between Class\,0 and Class\,I 
on the one hand, and Class\,II and Class\,III on the other. The higher X-ray temperature
measured for protostars could be a bias that results from obscuration of the soft X-ray
component (see references given above for more details). 

One problem of comparing X-ray properties of Class\,I protostars to
those of Class\,II and III pre-main sequence stars is the uncertainty
in the evolutionary stage itself. The standard classification scheme
uses \textit{Spitzer} photometry to distinguish between the different YSO classes.
This scheme is based on mid-IR colors or the slope of the mid-IR spectral energy distribution \citep[for example]{2004ApJS..154..363A, 2005ApJ...629..881H}.
Longer wavelength data is required to
uniquely identify the presence or absence of envelopes that are characteristic
of protostars \citep[e.g.][]{Motte98.0}.

Here we present the discovery of  an X-ray flare from the strongly embedded YSO \isooph 
in the $\rho$\,Ophiuchi star-forming complex, for which  we constrain the evolutionary stage from the analysis of its full SED. 
\isooph was firstly  detected at $1.3\,{\rm mm}$ by \cite{Motte98.0}. Emission at these wavelengths (that is ascribed to
an envelope) is a typical indicator for Class\,0 and Class\,I protostars, 
although YSO sample selection through $1.3\,{\rm mm}$ emission can also pick-up some 
Class\,II objects \citep[e.g.,][]{Enoch09.0}. The possible protostellar nature of
\isooph, combined with its X-ray flaring activity, have led us to examine this object 
in more detail.
A first characterization of \isooph was made by  \cite{Comeron93.1}, who detected it only in the $K$ band during
their near-IR survey. A tentative extinction of $A_{\rm V} \sim 45\,{\rm mag}$, a bolometric 
luminosity of $0.2\,L_\odot,$ and a mass of $0.3\,M_\odot$ were derived by a comparison with 
evolutionary models. \cite{Bontemps01.1} established ISO-Oph\,85
as a Class\,II object, based on the slope of its SED in the  $2-14\,{\rm \mu m}$ range. 
The \textit{Spitzer} measurements obtained within the c2d (\textit{From Molecular Cores to Planet-Forming Disk}) program  confirmed the 
presence of mid-IR excess \citep{Evans09.0}. 
The X-ray emission associated with \isooph 
was first reported in the 3XMM Catalogue \citep{2015arXiv150407051R}, 
and recognised as transient emission
within the framework of the project EXTraS ({\it Exploring the X-ray Transient and variable  
Sky},  \citealt{2015arXiv150301497D})\footnote{www.extras-fp7.eu} during a test of algorithms for 
the detection of transient phenomena.

The EXTraS project, aimed at the thorough characterization
of the variability of  X-ray sources in archival {\em XMM-Newton} data, is funded
within the EU Seventh Framework Programme for a data span of three years,
which started in January 2014. The EXTraS consortium is lead by INAF (Italy) and includes other five institutes 
in Italy, Germany, and the United Kingdom.

EXTraS will enable a significant step forward in the understanding
and characterization of X-ray variability, over time-scales ranging
from less than a second to up to several years (since the \textit{XMM-Newton} launch in 1999),
for different kinds of variability: short-term
(within a single observation), long-term (including analysis of slew
data between pointings), periodic modulations, and transients. EXTraS will 
produce the widest catalog of soft X-ray variable sources  to date and 
will characterize their variability via a series of data products.
The EXTraS database and tools will be released to the community at the end of the project.

\isooph was not detected in a 
systematic study of X-ray emission from YSOs in the same {\em XMM-Newton} 
data in which we find the transient emission, because source
detection was performed only on the time-averaged dataset \citep{Pillitteri10.1}.
This  shows the importance of time-resolved analysis procedures for identifying X-ray
properties of variable and absorbed YSOs, the potential of which
we illustrate here in the example of \isooph.

The structure of the paper is as follows.
In Sect.~\ref{sect:extras}, we describe the detection techniques and algorithms used within EXTraS.
In Sect. \ref{sect:xrays}, the X-ray time-resolved source detection is described. The X-ray data analysis for flare and quiescence is described in Sect. \ref{sect:xanalysis}. 
In Sect.~\ref{sect:sed} a fit of the IR Spectral Energy Distribution (SED) is performed using published and new multiwavelength photometry
to characterize the star and its circumstellar environment, including disk, envelope, and extinction.
The results from the SED analysis and the interpretation of
the X-ray emission are discussed in Sect.~\ref{sect:discussion}.

\section{Time-resolved source detection in EXTraS}\label{sect:extras}

The automatic transient search algorithm developed during the feasibility study phase 
and the initial stages of the EXTraS project is intended to detect transient sources in 
the data collected by all three {\it European Photon Imaging Camera} (EPIC) instruments, by performing a source detection on images 
accumulated over  time intervals that are much shorter than the total observation duration.
The pipeline is C-shell-based: it runs C-shell, C++, Python scripts and uses HEASOFT
6.15.1 FTOOLS and {\em XMM-Newton} Science Analysis Software (SAS) 13.5 tasks, all driven by a master C-shell script. It can operate on multiple observations, different 
instruments and combinations, operating modes, and energy bands. 
In its basic version, the search for transients is performed on time intervals of fixed duration, following 
these steps: data selection and standard cleaning; source detection performed with a maximum
likelihood algorithm with a SAS tool, \texttt{emldetect,}
over the whole observation in selected energy bands; source detection (same parameters
as above) on images with fixed time bins; comparison between the source list obtained for 
every bin and the source list referred to the whole observation.
We define `transient' candidates as all the sources detected in at least one time interval, but
with no counterpart in the latter list of the sources detected in the time-averaged image.
The 
transient search is affected by 
the soft proton background variability, the timescale of which is tens of seconds. Since this 
background rather uniformly affects the whole field of view, it is easily distinguished from 
the point source variability 
by the detection algorithm.


We also used a more effective detection technique, based on a modified version of the bayesian blocks algorithm. This adaptive-binning technique segments the 
observation in time intervals (blocks) in which the count rate is perceptibly constant, i.e., does not show statistically significant variations. 
Our modified version (mBB in the following) is described in Vianello et al. (in preparation), and can account for highly-variable background such as that 
found during proton flares in {\it XMM-Newton} data. In short, for each observation we divide the field of view into partially-overlapping 30" x 30" regions and we 
run mBB on each of them. Regions with no signal return only one block, which covers the whole observation, while regions containing candidate transients, 
return three or more blocks. We are assuming here that the transients we are looking for are much shorter than the duration of the observation. Among all
candidates found by mBB we consider as real astrophysical sources only those excesses that have a spatial distribution consistent with the point spread function (PSF) of a point source. 
False positives can also be detected by mBB because of particle events, 
bright pixels, and other instrument-related effects. This is verified by running the standard source-detection algorithm on the candidates provided by mBB.
\begin{figure} 
\begin{center}
\includegraphics[width=9.0cm]{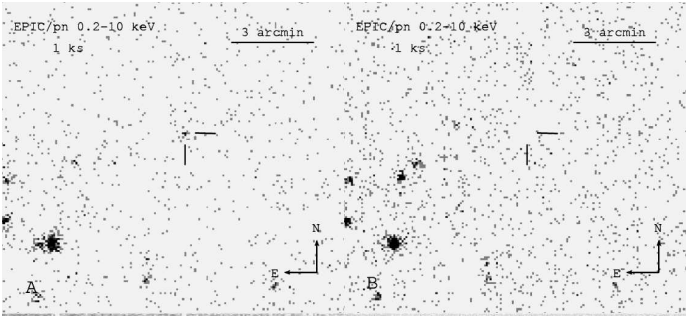}
\caption{Two EPIC PN exposures in the energy band $0.2-10.0\,{\rm keV}$, centered on the position of the X-ray detection associated with \isooph 
(lines, RA$=246$\textdegree$.742829\textdegree$, DEC=$-24$\textdegree$.628044\textdegree$), each one $1\,{\rm ks}$ long, taken few
thousands of seconds apart. The reference position for \isooph (\textit{HST}/NICMOS position from \citealt{Allen02.0}) is $2.2\,{\rm ''}$ away. These are the images extracted during the EXTraS feasibility study,
from which the X-ray emission was recognised as transient. 
In the left image, the source is visible, while in the right image, it is not. 
Other transients are visible in the left side of the field. This will be the subject of a future systematic study.
}
\label{fig:sed}
\end{center}
\end{figure}

\section{X-ray detection of ISO-Oph\,85}\label{sect:xrays}

One of the test fields for the transient source-detection algorithm developed
within EXTraS was the $\sim138$\,{\rm ks} long \textit{XMM-Newton} observation number 0305540701, 
 part of the {\em Deep Rho Ophiuchi X-ray observation} (DROXO), a $500\,{\rm ks}$
long {\em XMM-Newton} observation of core\,F in $\rho$\,Oph \citep{Sciortino06.1}.
The transient X-ray emission from a point source at a position consistent with \isooph was discovered in the data from the PN EPIC instrument,  both  using the pipeline with fixed time 
bins (bin time of $3000\, {\rm s}$) and using bayesian blocks analysis (in a time-interval of $\sim2500\,{\rm s}$ duration) 
in the $0.2-10.0\,{\rm keV}$ energy band. Using the pipeline-produced reference source list for the full observation without any soft proton filtering, 
\isooph is detected as a transient candidate, i.e., our pipeline does not detect it in the time-averaged image.

\begin{figure}[!htb]
\begin{center}
\includegraphics[width=9.5cm]{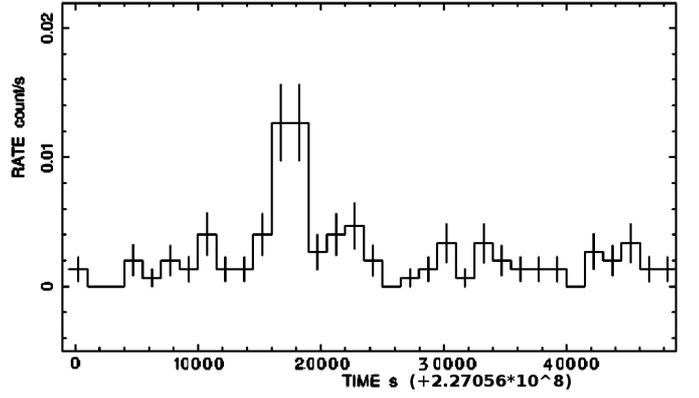}
\caption{X-ray lightcurve for the \isooph source region ($20''$ radius, centered on the X-ray detection position) 
for the $50\,{\rm ks}$ of GTI of the observation $0305540701$
in which the flare is detected. Bin width $1500\,{\rm s}$. Bars represent $1\,{\rm \sigma}$ uncertainties.}
\label{fig:lc}
\end{center}
\end{figure}

To  establish the duration of the flare exactly we run the \textit{XMM-Newton} SAS \texttt{edetect\_chain} 
 source detection routine on the PN events that were filtered for different time slices ($10000\,{\rm s}$, $1000\,{\rm s}$, $500\,{\rm s,}$ and $100\,{\rm s}$)
and in different energy bands ($0.3-2.0\,{\rm keV}$, $2.0-12.0\,{\rm keV}$, $0.3-12.0\,{\rm keV}$). 
We start with the longest interval ($10000\,{\rm s}$) and then proceed 
to successively shorter intervals when, in the longer one, a detection is present at
a position consistent with that of the transient source. 
The transient duration is $\sim2200\,{\rm s}$, during which the source is detected with a significance $>5\,{\rm \sigma}$ in the broad band $0.3-12.0\,{\rm keV}$. 
The distance between the X-ray source position and the \textit{HST}/NICMOS position of \isooph, which we will use as a reference position throughout this paper, is $2.2\,{\rm''}$.
Following \cite{2005A&A...431...87S}, we extract the number of detected X-ray sources with
USNO-B1 counterparts and calculate the probability of a random association between
the detected transient and \isooph with $P=1-e^{-(\pi r^2 \mu)} $, where $r$ is the 
X-ray source position uncertainty and $\mu \approx10^{-4} {\rm src/arcsec^2 }$, the numerical density of
X-ray sources. The chance probability is $<0.6{\rm \%}$, suggesting that the identified transient X-ray source
is associated with \isooph. 

The available \textit{XMM-Newton} observations of the $\rho$ Ophiuchi core\,F are all  
DROXO observations (0305540501, 0305540601, 0305540701, 0305540901, 0305541001, and 0305541101) plus a previous
$34\,{\rm ks}$-long observation 0111120201  \citep{2003A&A...408..581V}.
Observation 0305541001 was excluded from spectral analysis
because of the very high soft proton contamination. Adding up all the available observations, we have a grand total of $209.1\,{\rm ks}$ of 
observation (Good Time Intervals, GTI, free from soft proton background, obtained 
using the \texttt{bkgoptrate} SAS tool). We perform a search for other transients from \isooph with the automated pipeline and visually inspect the images 
but no other flares were detected. We perform an analysis of the X-ray emission, extracting 
the energy spectrum (for the flare) and the light curve (for the whole GTIs of the observation $0305540501$, in which the flare was detected, see Fig. \ref{fig:lc}). 

In all DROXO observations the source position falls
in a CCD gap  for both MOS1 and MOS2  instruments, and so we only consider PN camera data. 
For each observation, the background is extracted 
for the flare analysis from a circular source-free region 
on the same CCD as the source. During the flare, we observe a net source count rate of
$1.762\cdot10^{-2}\pm3.269\cdot10^{-3}\,{\rm cts/s}$ in the band $0.3-12.0\,{\rm keV}$.
An X-ray luminosity upper limit is obtained for the quiescent regime of the source  outside the flare, in which the source is not detected in X-rays within $3{\rm \sigma}$ by \texttt{edetect\_chain}. 
 We consider all the GTIs of all the observations except the $2200\,{\rm s}$ of the flare as the quiescent time.  
For the quiescence, we obtain the background subtracted count rate upper limit at $3{\rm \sigma}$ from the formula for the signal-to-noise ratio
\begin{equation}
\frac{S}{N}=\frac{R_{src}\cdot T_{exp}}{\sqrt{R_{bkg}\cdot T_{exp}\cdot \frac{A_{src}}{A_{bkg}}+R_{src}\cdot T_{exp}}}
,\end{equation}
where the signal-to-noise ratio (S/N) is in $\sigma$ and set $(S/N)=3$, $R_{src}$ is the source count rate upper limit, 
$R_{bkg}$ is the background count rate, $T_{exp}$ is the total exposure time minus the flaring time ($206.9\,{\rm ks}$), and $A$ is the area of the extraction region.
The background count rate is obtained from a background extraction region of area  $\sim7$ times the source region on the same CCD of the source extraction region.
For the whole energy band ($0.3-12.0\,{\rm keV}$) we obtain a count rate upper limit of $4.91\cdot10^{-4}\,{\rm cts/s}$.

\section{X-ray properties of \isooph}\label{sect:xanalysis}

\subsection{Flare light curve}

The X-ray light curve (Fig. \ref{fig:lc}) shows a clear count excess in two $1500\,{\rm s}$-long adjacent time bins, corresponding to the detected flare.
Because of the few counts ($39$ events from the 
whole $2200\,{\rm s}$ flaring time), it is not possible to infer  the shape nor a decay time for the flare. We can only state that
the decay time is less than $3000\,{\rm s}$ (twice the bin size).
The low number of counts  prohibits a standard spectral analysis.  We observe a significant counts excess 
in the range $2.0-6.0\,{\rm keV}$ with respect to a pre-flare and post-flare $20\,{\rm ks}$ interval (including mainly background counts), which suggests an impulsive heating episode, as  is typical for a flare.

\subsection{Flare spectral analysis}
A technique for performing a spectral analysis on few-counts spectra, using quantile quantities,
was developed by \cite{2004ApJ...614..508H} and used for the analysis of YSOs spectral
properties (e.g. \citealt{2007A&A...468..405S}).  We calculate the $25{\rm\%}$, $50{\rm\%,}$ and $75{\rm\%}$ 
quantiles  of the counts of the   flare observed from \isooph. 
 Then we derive the position of \isooph in the quantile space examined by \cite{2004ApJ...614..508H},
defined by $Q_{25}/Q_{75}$ vs $Q_{50}/(1-Q_{50})$. 

A $\log T$ vs $N_{\rm H}$ theoretical grid, predicted 
by an absorbed one-component coronal thermal model (WABS*APEC) with fixed abundance ($Z = 0.3 Z_{\odot}$ from
\citealt{1989AIPC..183....1G}), and modeled via $100000\,$\textit{s} simulated spectra for a grid with 
$\log{T}$ from $6.50$ to $8.00$ 
and $N_{\rm H}$ from $1\cdotp10^{20}\,{\rm cm^{-2}}$ to  $N_{\rm H}=1\cdotp10^{24}\,{\rm cm^{-2}}$, 
can be superimposed onto the phase space, thus giving the possibility to  determine $kT$ and $N_{\rm H}$ 
values for each point in the quantile space (Fig.\ref{fig:sed}). Following \cite{2007A&A...468..405S}, we compute this diagram for EPIC/PN data in the energy band $0.5-7.3\,{\rm keV}$.
We extract a spectrum  of the $2200\,{\rm s}$ total flare interval during which the source
was detected, for a region with a radius of $20{\rm ''}$ that is centered on the \textit{Hubble Space Telescope} \textit{HST}/NICMOS
position of \isooph, which has the highest precision. We extract a background spectrum from a region on the same detector CCD.
From the position of \isooph on the quantile grid, we obtain 
$kT=1.15_{-0.65}^{+2.35}\,{\rm keV}$ and $N_{\rm H}=1.0_{-0.5}^{+1.2}\cdot10^{23}\,{\rm cm^{-2}}$. 
The quantiles of \isooph represent the median value of the distribution that results  when the quantile calculation is executed
$1000$ times with randomly selected background photons according to the area-scaling factor between the source and background extraction regions.
The error on the $x$ axis is derived from the statistical error of \cite{2004ApJ...614..508H}. 
For each of the $1000$ quantile calculations, we have a statistical error. The distribution of these errors shows a peak at $0.075$. 
We take this value as our error. 
On the $y$ axis, this error is overestimated (see \citealt{2004ApJ...614..508H}). Thus, we do a simple error propagation on $Q_{25}/Q_{75}$ for each of the 
$1000$ calculations.
The distribution of these errors shows a peak at $0.12$, which we assume to be our error on the $y$ axis.

To put the spectral properties of \isooph in context to those of the other
YSOs in $\rho$\,Oph, we calculate $Q_{25}/Q_{75}$ and $\log{(Q_{50}/(1-Q_{50}))}$ quantiles for all 
DROXO X-ray sources from \cite{Pillitteri10.1} using the original extraction regions for
source and background events of the full GTI filtered observing time. 
The background regions are larger than the source
regions, typically by a factor of $\sim3$. By way of an analogy to the case of \isooph, and taking into account this scaling factor, 
we randomly select the appropriate fraction of photons from the background events 
files and  perform the quantile determination for each DROXO source $1000$ times, 
each time with a different ensemble of background photons. 
The values shown in Fig. \ref{fig:sed}
represent the medians of the resulting distributions for $Q_{25}/Q_{75}$ and $\log (Q_{50}/(1-Q_{50}))$.

We classify the DROXO sources as Class\,I, flat, Class\,II, and Class\,III 
YSOs, according to their SED slope ($\alpha_{\rm SED}$)
as defined and computed by \cite{Evans09.0}. Each YSO class is represented by a different
plotting symbol in Fig. \ref{fig:sed}. 
Using the same definition of $\alpha_{\rm SED}$, in Sect. \ref{sect:sed} we 
identify ISO-Oph\,85 as a flat source. 
As can be seen from Fig. \ref{fig:sed}, the position of ISO-Oph\,85 in the quantile phase space 
is consistent with that of Class\,I, ``flat'', but also Class\,II sources in $\rho$\,Oph, because of the large error bars.
We note that the point of \isooph in the diagram is only due to the detected flare emission, while the DROXO stars are shown at their time-averaged spectral parameters.


\begin{figure}[!htb]
\begin{center}
\includegraphics[width=10.0cm]{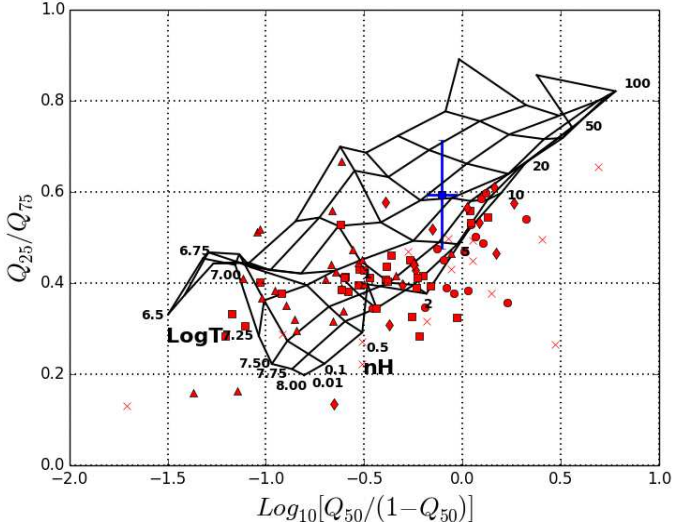}
\caption{Phase space for the quantile analysis, with the parametric grid $\log T$, $N_H (10^{22}\,cm^{-2})$ superimposed. The flare on \isooph is
represented by the blue square. DROXO time-averaged  sources are represented by the following symbols: Class\,I (circles), Flat (diamonds), Class\,II (squares), Class\,III (triangles), undefined (x).
}
\label{fig:sed} 
\end{center}
\end{figure}

We then evaluate the X-ray flare emission properties with XSPEC \footnote{XSPEC (X-Ray/Gamma-Ray Spectral Analysis Package) is a 
package for X-ray and Gamma-ray spectral analysis, provided by HEASARC NASA \citep{1996ASPC..101...17A}}.  
Within XSPEC we use a {\sc WABS $\cdot$ APEC} model  
with an X-ray temperature of $kT = 1.15\,{\rm keV}$ and a hydrogen column density $N_H=10^{23}\,{\rm cm^{-2}}$, as obtained from the quantile analysis, 
to get the $0.3-12\,{\rm keV}$ extinction-corrected X-ray flux 
during the flare ($f_{\rm x} = 6.78 \cdot 10^{-12}\,{\rm erg\,cm^{-2}\,{\rm s}^{-1}}$),   
the X-ray luminosity ($\log{L_{\rm x}\,[{\rm erg\,{\rm s}^{-1}}]}\,= 31.6^{+2.0}_{-1.2}$) and the activity index 
($\log{(L_{\rm x}/L_{\rm *,SB})} = -2^{+2}_{-1}$). The bolometric luminosity , $\log{L_{\rm *,SB}\,[{\rm erg\,{\rm s}^{-1}}]}=33.1^{+0.2}_{-0.1}$ is that obtained with the Stefan-Boltzmann law using the effective temperature and stellar radius of the best-fit model.


\subsection{X-ray luminosity during quiescence}

From the upper limit to the quiescent count rate (as given  in Sect. \ref{sect:xrays}) times a conversion factor, we obtain an X-ray flux and luminosity upper limit. 
 We calculate the conversion factor from the 
flux obtained with XSPEC for a WABS$\cdot$ APEC model, using for $kT$ and $N_H$ the values from the quantile analysis. 
Given the distance of the source ($120\,pc$, \citealt{2008A&A...489..143L}), we obtain the luminosity upper limit,
that is $\log{L_{\rm x}}<29.5 $.  We note that the formally most 
conservative upper limit,  which we can derive from the uncertainties of $N_H$ and $kT$ obtained from the quantile diagram, amounts to $\log{L_{\rm x}}<31.6 $.      
By increasing the temperature, this luminosity decreases.
\cite{Pillitteri10.1} gives an upper limit of $\log{L_{\rm x}}<28.54$. This value  was obtained with a different 
energy band ($0.3-10.0\,{\rm keV}$), model APEC component temperature ($kT=3.1\,{\rm keV}$), exposure time ($300.9\,{\rm ks}$), and a different absorption $N_H=2.3\cdot 10^{22}\, {\rm cm^{-2}}$. 
Moreover \cite{Pillitteri10.1} calculate the upper limits using a different technique, following \cite{1997ApJ...483..350D}. 
This makes it difficult to compare the results. Nevertheless, $kT$ and $N_H$ calculated by \cite{Pillitteri10.1} are consistent with the error bars of the \isooph point in Fig. \ref{fig:sed}.


\section{Spectral energy distribution}\label{sect:sed}

The SED is  key to  understanding  the evolutionary phase of a YSO.
To this end, we collect all available photometric data for ISO-Oph85 from the literature and from data archives, 
we reanalyse Spitzer photometry, and we present as yet unpublished \textit{Herschel} photometry of ISO-Oph85.

\begin{figure}[!htb]
\begin{center}
\includegraphics[width=9.5cm]{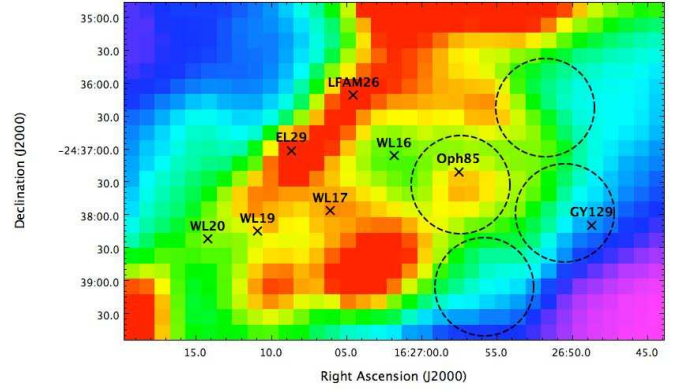}
\caption{The {\em Herschel}/SPIRE\,$500\,{\rm \mu m}$  field with the position of \isooph and the other millimetric sources in the field of view \citep{Motte98.0}. The source extraction region is shown,  centered on the emission peak in SPIRE\,$500\,{\rm \mu m}$, together with the three circular regions used for the background evaluation, chosen to avoid
contamination from other known bright YSOs and dust filaments.}
\label{fig:spire}
\end{center}
\end{figure}

\begin{figure*}[!htb]
\begin{center}
\parbox{18cm}{
\parbox{18cm}{
\includegraphics[width=18cm]{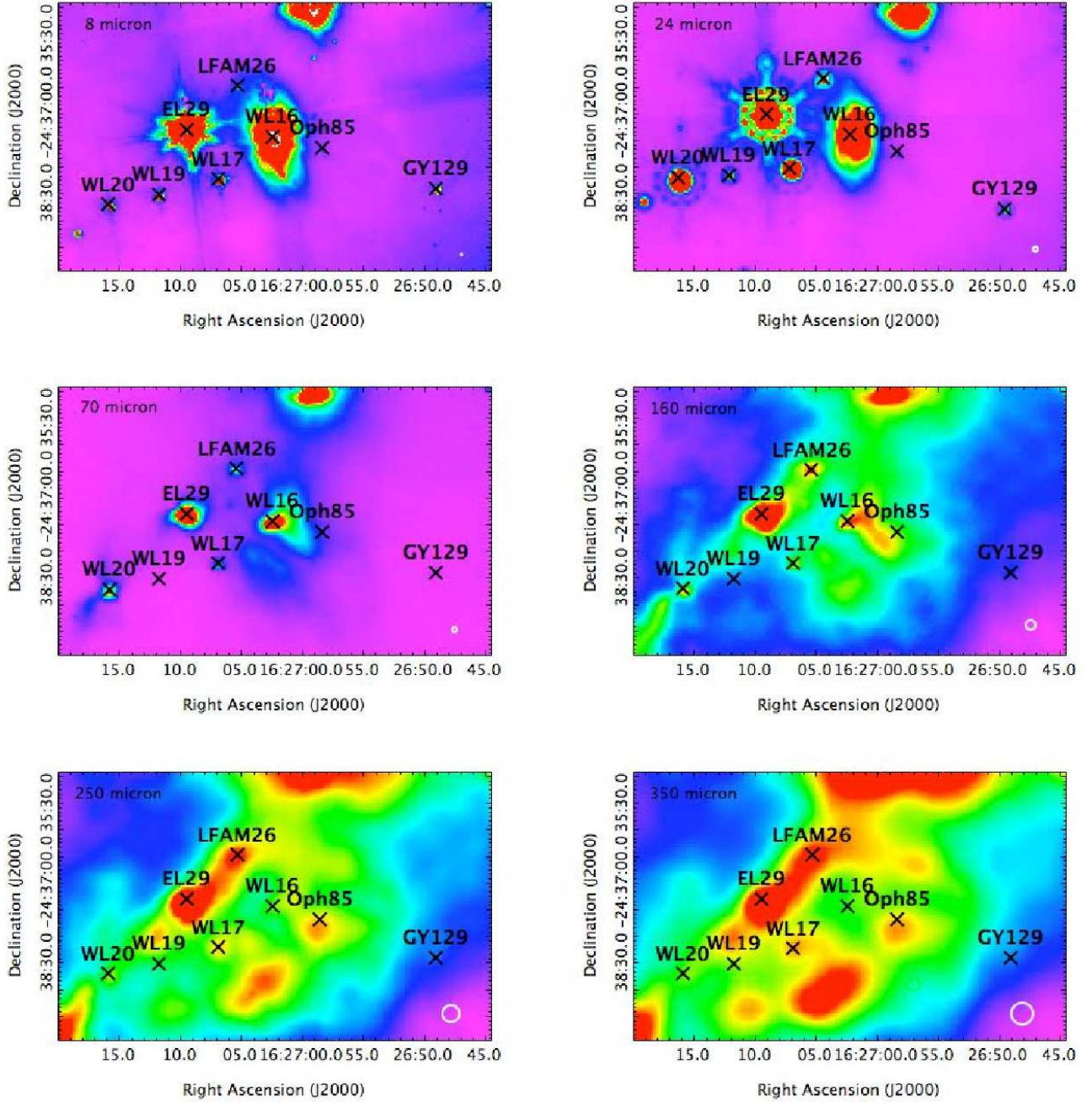}
}
}
\caption{Images from \textit{Spitzer} IRAC $8\, {\rm \mu m}$, MIPS $24\,{\rm \mu m}$, {\em Herschel} PACS $70\,{\rm \mu m}$, PACS $160\,{\rm \mu m,}$ and SPIRE $250\,{\rm \mu m}$  and $350\,{\rm \mu m}$ instruments
with the \textit{HST}/NICMOS position of \isooph and the position of the millimetric sources in the field of view \citep{Motte98.0}. The instumental PSF is shown in the 
bottom right of each panel.}
\label{fig:pacs}
\end{center}
\end{figure*}

\subsection{Multi-band photometry for ISO-Oph85}
We cannot find any report of optical photometry for this object, presumably 
because of its large extinction. 
In some near-IR filters only the upper limits have been measured while mid-IR 
detections are available from various space missions 
({\em Spitzer},  \textit{WISE}, and ISOCAM).  \textit{WISE}  data in the W3 and W4 bands are listed in 
Table~\ref{tab:photometry} but are 
not considered further because they carry a source contamination flag and, in fact, no source is
seen on the images of W3 and W4 bands. The photometric point with the 
longest wavelength is the $1.3\,{\rm mm}$ flux measured with the IRAM $30\,{\rm m}$ telescope by 
\cite{Motte98.0}. 

The \textit{Spitzer} data are taken from the \textit{Cores to Disks Legacy Project, (c2d, PI Evans, pid 177)}
We perform aperture photometry in the \textit{Spitzer} IRAC bands ($3.6, 4.5, 5.6,$ and $8\, {\rm \mu m}$) 
by using the products available from the {\em Spitzer Science Center} (SSC) archive, and by 
adopting the recommended apertures{\footnote{IRAC Instrument
Handbook}} at all wavelengths. 
These, for a reference pixel size of $1.2\,{\rm ''}$, correspond to $12\,{\rm''}$ (source aperture) and to $14.4\,{\rm''}$ and $24\,{\rm ''}$ 
(sky apertures). This way we obtain flux densities that agree within  $5\,{\rm \%}$ with the values 
provided by the \textit{c2d} collaboration \citep{Evans09.0}.
We also perform aperture photometry in the MIPS 24 ${\rm \mu m}$ band using a $13{\rm ''}$ source aperture and 
$20{\rm ''}$ to $32{\rm''}$ sky apertures{\footnote{MIPS Instrument Handbook}}. In this case, the computed flux is $\sim 40\, {\rm \%}$ 
higher than the value found by \cite{Evans09.0}. We investigate the source of a such a discrepancy. To this 
end, we notice that the c2d collaboration carried out their photometric measurements using a point-source profile-fitting approach. Along these lines, we perform PSF-fitting of our source in the 
Mopex/Apex{\footnote{http://irsa.ipac.caltech.edu/data/SPITZER/docs/dataanalysistools
/tools/mopex/}} environment, where, as 
an input PSF, we use the MIPS 24 $\mu m$ template that is publicly distributed by the SSC. After optimization of the Apex parameters, we 
obtain a flux-density value consistent with \cite{Evans09.0}, although with a poor reduced $\chi^2$ ($9.3$). 
Accordingly, and for consistency with the IRAC measurements, 
we decide to adopt the higher value derived from our aperture photometry.
\textit{Spitzer} MIPS  $70\,{\rm \mu m}$ data also exist, but the quality of these data is typically rather 
poor compared to data at the same wavelength from \textit{Herschel} PACS $70\,{\rm \mu m}$ since they, as we
ascertain, show flux non-linearities and artifacts.
This is the result of the different types of detectors. We thus decide not to use them.

We also analyse as yet unpublished submillimeter photometry 
from the \textit{Herschel} mission (Fig. \ref{fig:spire}).
For the position of  \isooph, archival \textit{Herschel} PACS and SPIRE parallel mode observations at $70\,{\rm\mu m}$, 
$160\,{\rm \mu m}$, $250\,{\rm \mu m}$, $350\,{\rm \mu m}$, and $500\,{\rm \mu m}$ from the 
``Probing the origin of the initial mass function: A wide-field \textit{Herschel} photometric survey of nearby star-forming cloud 
complexes.'' program (KPGT\_pandre\_1) are available. The data have an angular resolution ranging from $6\,{\rm ''}$ (at $70\,{\rm\mu m}$) 
to $35\,{\rm ''}$ (at $500\,{\rm \mu m}$). 

Given the difference in angular resolution between the NIR and mid-IR/submm data, it is important to make sure 
that the source that is visible in the lowest resolution band, i.e., \textit{Herschel}/SPIRE $500\,{\rm \mu m}$, is indeed the same source that appears 
at shorter wavelengths. To this end, we adopt the location of 
the \textit{HST}/NICMOS peak of emission as a reference. We then verify that the position of 
the peaks of emissions in each of the \textit{Herschel}/PACS and SPIRE bands are within a FWHM from the position of the NICMOS peak, where 
the FWHM defines the PACS or SPIRE instrument resolution at a given wavelength. 
Fig. \ref{fig:spire} shows an offset between the position of \isooph in \textit{HST}/NICMOS and in \textit{Herschel} SPIRE $500\,{\rm \mu m}$, 
but it is well within the angular resolution of SPIRE $500\,{\rm \mu m}$. Besides, there are no other millimetric sources that can be associated with the 
position of the emission peak in  SPIRE $500\,{\rm \mu m}$, and generate confusion in the association with \isooph.  We, therefore, 
associate the SPIRE emission peak with ISO-Oph85, but caution that the positional offset between 
submm and IR is larger than for most other YSOs in the region. 

We reduce \textit{Herschel} data  via the \textit{Herschel} reduction pipeline.
We perform aperture photometry in the PACS and SPIRE 
bands. At IR and submm wavelengths, the background emission surrounding \isooph is 
highly structured and position-dependent (see Fig. \ref{fig:pacs}). For this reason, in each band we derive the source flux 
by taking the average of the values obtained by estimating the background from sky apertures that were placed 
at different locations in the source proximity. Using the median, the results do not change. Accordingly, the dispersion of the flux measurements 
around the mean value, combined with the intrinsic statistical errors, provide the flux errors. We use source and sky apertures of $12\,{\rm ''}$, $22\,{\rm ''}$, $40\,{\rm ''}$ 
$50\,{\rm''}$, and $60{\rm ''}$ at $70\, {\rm \mu m}$, $160\, {\rm \mu m}$, $250\,{\rm \mu m}$, $350\,{\rm \mu m}$, and $500\,{\rm \mu m}$. In the PACS bands, \isooph is barely visible above the background, therefore at $70\,{\rm\mu m}$ and  $160\,{\rm \mu m}$
we are only able to place  flux upper limits.

\subsection{SED analysis}
We model the SED of \isooph using the online SED fitter developed by 
\cite{Robitaille07.1}. This algorithm compares the observed photometric data 
set with a precomputed grid of $200 000$ theoretical SEDs, which comprises $14$ free
parameters that characterize the star, the disk, and the envelope of the hypothetical YSO. 
As pointed out by \cite{Robitaille07.1}, because of the complexity of the $\chi^2$ surface, 
this approach does not allow  the true parameters of a YSO to be strictly determined, but,
depending on the quality of the observed SED, it
does allow meaningful constraints to be placed on a number of these parameters. 
Additional parameters are the interstellar extinction ($A_{\rm V}$) and the distance ($d$). 
Throughout the fitting process both $A_{\rm V}$ and $d$ are
allowed to vary in a range that we set to $A_{\rm V} = 0 - 80\,{\rm mag}$ and
$d = 120 - 130\,{\rm pc}$. These choices are motivated by the previous extinction estimate
of \cite{Comeron93.1} and the mean and spread of the distances cited in the literature
for $\rho$\,Oph \citep[see][for a summary]{Mamajek08.0}.

\begin{figure}[!htb]
\begin{center}
\includegraphics[width=8.5cm]{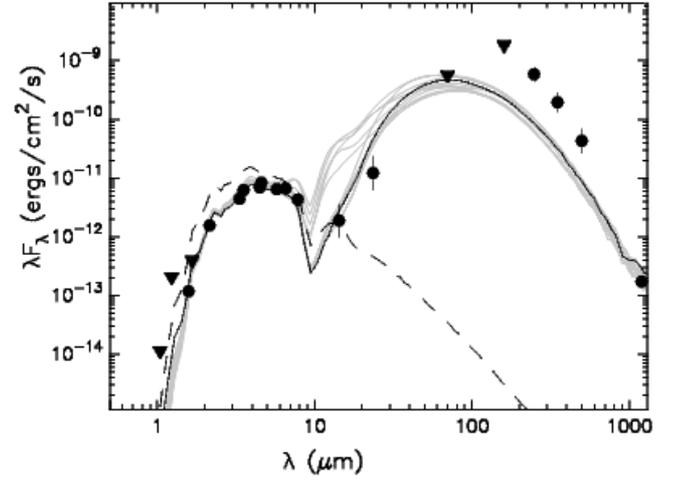}
\caption{Spectral energy distribution of \isooph and of the $37$ best-model fits. Upper limits 
are represented as downward pointing triangles. The best-fit model is shown as solid black 
line and the other `statistically acceptable' models as solid grey lines. The dashed line
indicates the photospheric contribution of the best-fit model.}
\label{fig:sedd}
\end{center}
\end{figure}

The filter list provided for  Robitaille's SED-fitting tool does not include 
the \textit{HST}/NICMOS and the ISOCAM filters. Therefore,   
we have added these data points to the input of the SED fitter 
using the ``monochromatic'' option as an approximation.
Given the uncertainties of the photometry related to zero-points and variability,
the error introduced by neglecting the shape and width of the filter transmission is
likely irrelevant. Following \cite{Robitaille07.1}, we impose a minimum of $10\,{\rm \%}$
flux error on all data points, 
and we base our assessment of the YSO parameters on the models with the smallest
$\chi^2$ values, i.e., the $37$ models with $\chi^2 - \chi^2_{\rm best} < 3$ where $\chi^2_{\rm best}$
represents the best fit model ($20$ degrees of freedom). These are shown together with the observed SED
in Fig.~\ref{fig:sedd}. 

As can be seen in Fig.~\ref{fig:sedd}, 
a long-wavelength hump (representing emission
from the envelope) is clearly visible. The hump is determined mainly by the \textit{Herschel} data. 
 If the hypothesis that the association with the {\em Herschel} peak is correct, this provides strong evidence for the protostellar nature of \isooph.
However, in this region of the SED the fit is not good. The uncertainties of the \textit{Herschel} data are notoriously difficult
to evaluate as a result of the strong spatial structure of the background emission at these wavelengths.  

\begin{figure*}
\begin{center}
\parbox{18cm}{
\parbox{6cm}{
\includegraphics[width=6cm]{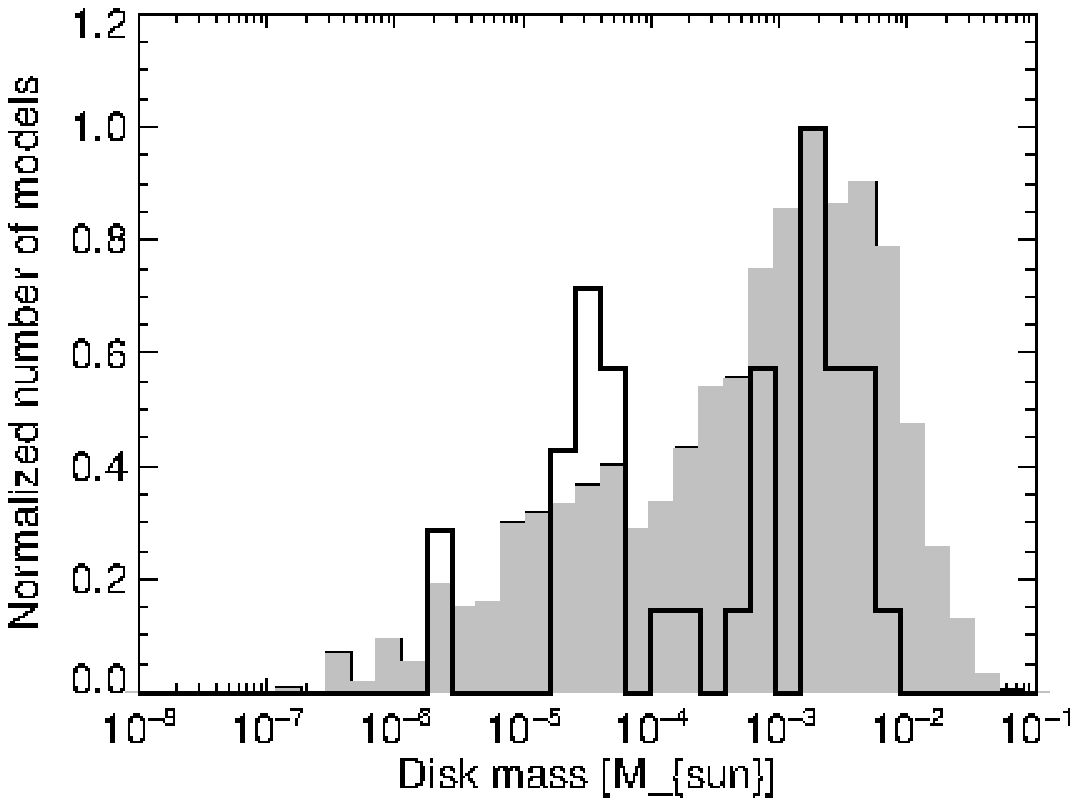}
}
\parbox{6cm}{
\includegraphics[width=6cm]{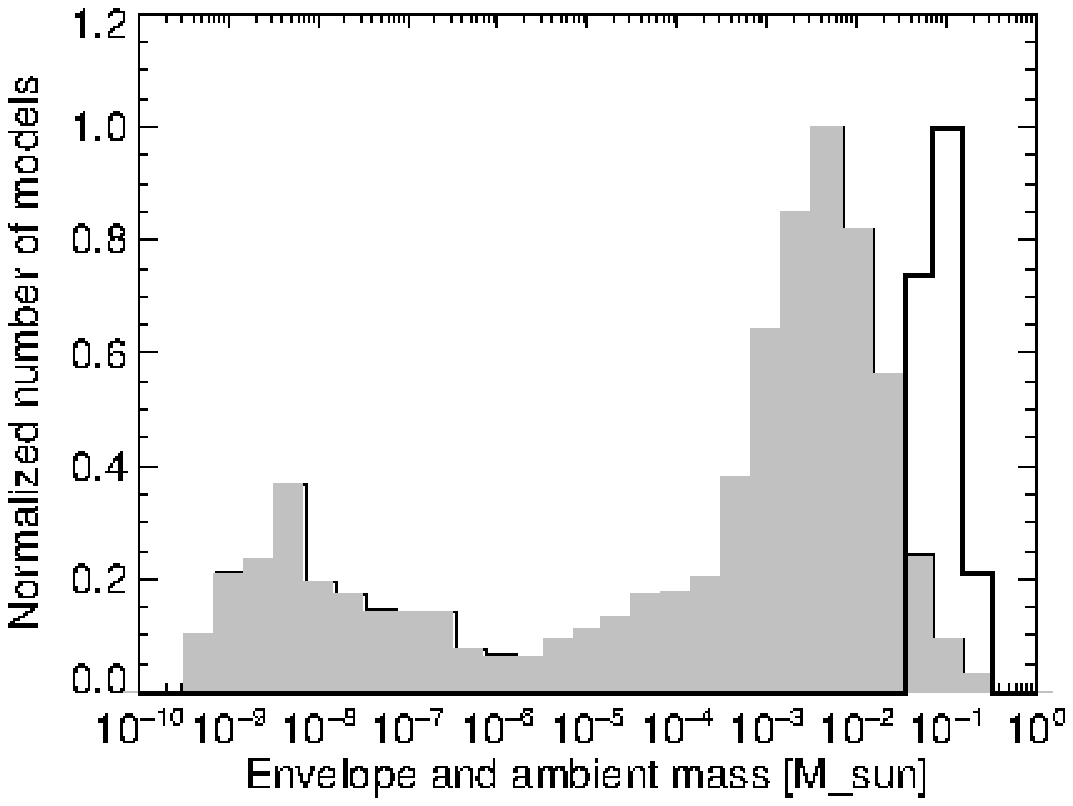}
}
\parbox{6cm}{
\includegraphics[width=6cm]{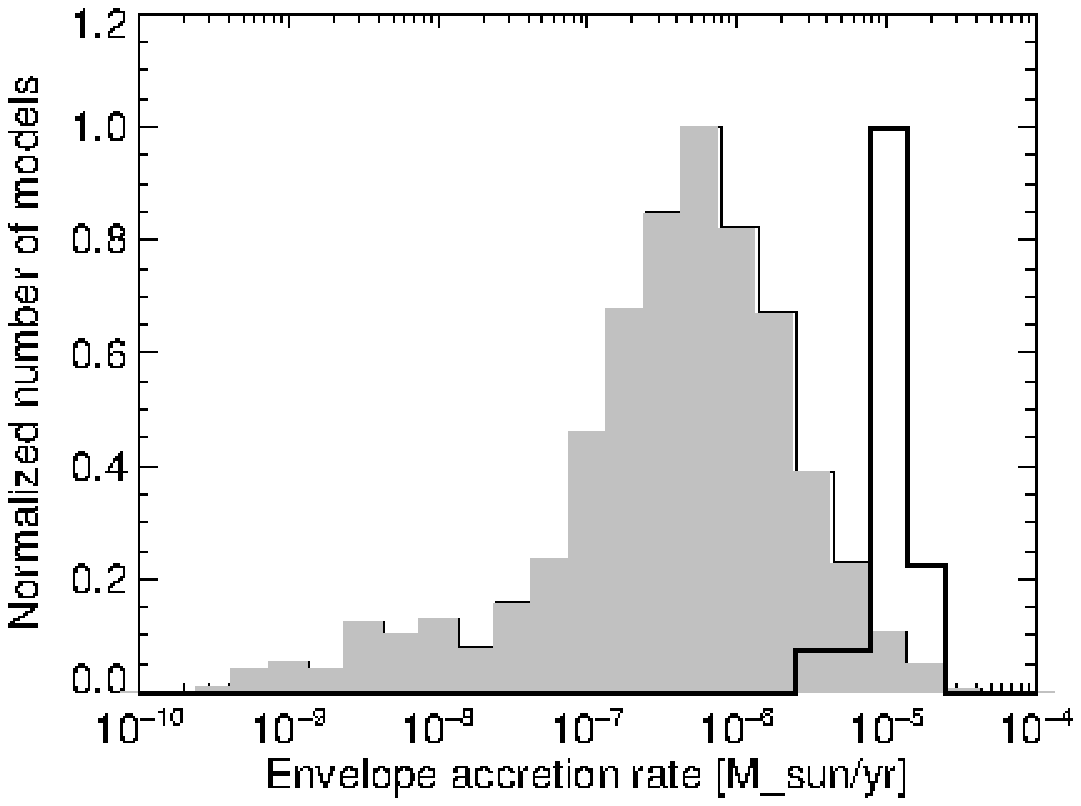}
}
}

\caption{Distribution of model parameters for the $37$ best-fit models (black histograms)
and for the $10 000$ best-fitting models (grey histograms). The distributions are normalized
to their peaks. From left to right: disk mass, envelope mass, envelope accretion rate.}
\label{fig:histograms}
\end{center}
\end{figure*}

The most evident feature in the distribution of the bestfit parameters 
is related to the masses of the disk ($M_{d}$) and  the envelope ($M_{\rm env}$). In Fig. \ref{fig:histograms}
we compare the disk and envelope masses for the $37$ best-fit models to those of the $10 000$ best fits.
This shows that all best-fit models are among those from the available models with the highest envelope mass and accretion rate. The range of acceptable disk
masses is large, while all bestfit models have high envelope mass.
The range of best fit parameters is given in 
Table~\ref{tab:sedparams}, where we also list the median and $25\,{\rm \%}$ and $75\,{\rm \%}$ quantiles 
for all fit parameters.

\begin{table}
\begin{center}
\tiny{\caption{Photometry for \isooph}
\label{tab:photometry}
\begin{tabularx}{\columnwidth}{llrrcrl}\hline
Instrument & Filter &   Mag &  Emag &  Flux &  Error &  Ref \\ 
&&&&density &&\\
           &        & [mag] & [mag] & [mJy] & [mJy]  & \\ \hline
\textit{HST}/NICMOS3 & $F110W$ &   21.5 &   --- &    0.004 &  --- &   (1) \\
2MASS & $J$ &       18.17 &  --- &    0.086 &  --- &   (2) \\
\textit{HST}/NICMOS3 & $F160W$ &   18.05 &  0.1 &    0.064 &  0.006 & (1) \\
2MASS & $H$ &       16.74 &  --- &    0.210 &  --- &   (2) \\
2MASS & $K_s$ &      14.41 &  0.08 &   1.145 &  0.084 & (2) \\
WISE  & $W$1 &        11.98 &  0.02 &   4.94 &   0.09 &  (3) \\
\textit{Spitzer}/IRAC & $I1$ &      &   &  7.5  &  0.75  &   \\
\textit{Spitzer}/IRAC & $I2$ &      &   &  10.3  &1.03 &   \\
WISE  & $W2$ &        10.31 &  0.02 &  12.80 &   0.24 &  (3) \\
\textit{Spitzer}/IRAC & $I3$ &      &   &  12.5 &   1.25 &   \\
\textit{ISO}/ISOCAM & $LW2$ &          9.37 &  0.1 &   15.00 &   4.00 &  (5) \\
\textit{Spitzer}/IRAC & $I4$ &       &   &  11.8 &   3.2 &   \\
WISE  & $W3^{(c)}$ &         8.17 &  --- &   15.64 &   --- &   (3) \\
\textit{ISO}/ISOCAM & $LW3$ &          8.10 &  0.10 &  11.00 &   7.00 &  (5) \\
WISE  & $W4^{(c)}$ &         4.78 &  --- &  101.27 &   --- &   (3) \\
\textit{Spitzer}/MIPS & $24\,\mu$m &     &   &  118 &   77 &  \\
\textit{Herschel}/PACS & $70\,\mu$m & &         &   13320&---&\\
\textit{Herschel}/PACS & $160\,\mu$m &&&         98890   &---&\\ 
\textit{Herschel}/SPIRE & $250\,\mu$m &&&                50610 &  14320           &\\
\textit{Herschel}/SPIRE & $350\,\mu$m  & & &              24110  &       8939                    & \\
\textit{Herschel}/SPIRE & $500\,\mu$m &&&           24110          &     3730               &\\
IRAM & $1.3mm$ &             &       &  70.00 &  15.0 &   (6) \\
\hline
\multicolumn{7}{l}{(c) - confusion flag} \\
\multicolumn{7}{l}{(1) - \protect\citep{Allen02.0}, (2) - \protect\citep{Cutri03.1}, (3) - \protect\citep{2008SPIE.7016E..18B},} \\
\multicolumn{7}{l}{(4) - \protect\citep{Evans09.0}, (5) - \protect\citep{Bontemps01.1},}\\
\multicolumn{7}{l}{(6) - \protect\citep{Andrews07.0}.} \\
\end{tabularx}
}
\end{center}
\end{table}

 

%

Using the definitions of
\cite{Robitaille07.1}, based on disk mass and envelope accretion rate, we can assign an
evolutionary `stage' to the  best-fit models. 
Stage\,0/I represents objects with significant infalling envelopes and possible disks,
and Stage\,II denotes objects with optically thick disks and possible remains of an envelope. 
All best-fit models for the SED of ISO-Oph85 correspond to Stage 0/I, underpinning the early 
evolutionary stage of the object.

%
\begin{table}\begin{center}
\caption{SED best-fit parameters. Errors represent $25\%$ and $75\%$ quantiles for the 37 statistically acceptable models as defined in Sect. \ref{sect:sed}.
($A_{V,fit}$ - total extinction, $A_{V,cs}$ -  circumstellar extinction, inside the envelope cavity)} 
\label{tab:sedparams}
\begin{tabular}{lcccc}\\ \hline
Parameter & ${\rm All bestfit}$  \\
$N_{\rm fit}$ & $37$ \\ \hline
                    $d [pc]$ & $      129.^{ 129.}_{ 123. }$ \\
           $A_{V,fit} [mag]$ & $       45.^{  46.}_{  42. }$ \\
            $A_{V,cs} [mag]$ & $        0.^{   5.}_{   0. }$ \\
   $M_{\rm d}* 10^3 [M_\odot]$ & $      0.81^{ 2.79}_{ 0.05 }$ \\
$\dot{M}_{\rm d} [M_\odot/yr]$ & $      -9.3^{ -9.8}_{ -8.6 }$ \\
         $R_{\rm d,in} [AU]$ & $     6^{   30}_{    1 }$ \\
        $R_{\rm d,out} [AU]$ & $          60^{  493}_{   42 }$ \\
     $M_{\rm env} [M_\odot]$ & $      0.12^{  0.17}_{  0.09 }$ \\
  $\dot{M}_{\rm env} [M_\odot/yr]$ & $     -4.91^{ -4.76}_{ -4.99 }$ \\
       $R_{\rm env,in} [AU]$ & $   142^{  784}_{   42 }$ \\
      $R_{\rm env,out} [AU]$ & $  1515^{ 1764}_{ 1395 }$ \\
           $T_{\rm eff} [K]$ & $  2960^{ 2986}_{ 2851 }$ \\
       $R_{\rm *} [R_\odot]$ & $      2.33^{2.56}_{2.18 }$ \\
               $M_{\rm *} [M_\odot]$ & $      0.14^{0.15}_{0.12 }$ \\
                 $Age [Myr]$ & $      0.12^{0.15}_{0.09 }$ \\
\hline
\end{tabular}
\end{center}
\end{table}

In Sect.\ref{sect:discussion} we compare ISO-Oph\,85 to other X-ray sources in $\rho$\,Oph
for which we use the YSO classes given by \cite{Evans09.0} that are based on the slope
of their SEDs ($\alpha_{\rm SED}$) from $2 - 24\,{\rm \mu m}$.  
For \isooph \cite{Evans09.0} determined $\alpha_{\rm SED} = -0.29$ using
2\,MASS and {\em Spitzer} IRAC and MIPS\,1 data. 
In our SED, we include analysis additional photometry from \textit{HST},  \textit{WISE,}  and ISO. The SED slope we find from a least-squares fit to the data  given in 
Table~\ref{tab:photometry}
in the $2-24\,{\rm \mu m}$ interval is $\alpha_{\rm SED} = 0.10$. 
However, for consistency with the classification of the other DROXO sources, we 
also use  the SED classification from \cite{Evans09.0} for \isooph .
According to these results, ISO-Oph85 is considered a `flat spectrum' source.
We note, however, that the spectral index of \isooph is dominated by the low $K$-band
flux and  would be much steeper if the $K$-band were removed. 

To summarize, our analysis of the SED of ISO-Oph85 establishes this object
as a bona fide Class I protostar. The \textit{Herschel} photometry  gives credibility to this object being a Class I protostar. 
A slight doubt remains with this classification because of the offset between the optical/IR and submm position of the emission peaks.
Given the large uncertainties in the \textit{Herschel} fluxes, we do not lend too much weight to the values
for the disk and envelope parameters derived from the SED fits.
The stellar parameters are likely robust against the uncertainties in the submm fluxes but subject
to uncertainties from the extinction.

\section{Discussion and conclusions}\label{sect:discussion}
The observation of a flare from \isooph is of twofold interest. Firstly, it provides a first validation of the discovery and 
science potential of the EXTraS project. Secondly, it gives insight into the physics, structure, and evolution
of the YSOs. We showed that the analysis of the whole SED is fundamental for recognizing the evolutionary stage of YSOs, while the identification 
of the YSO class without submm data in the SED can lead to erroneous classification.
This work also shows the possibility to perform a refined time-resolved analysis of the X-ray emission down to timescales that were
previously inaccessible, and how such an analysis is fundamental for the understanding of the properties of the X-ray emission from YSOs.
The object
\isooph was detected in millimetric wavelength by \cite{Motte98.0}. Such long wavelength emissions can be a sign of  protostellar nature, so the 
X-ray flaring activity from \isooph is particularly interesting, since X-ray flaring activity has rarely been observed in protostars.
However, the wide gap in the SED between mid-IR (\textit{ISO}/ISOCAM, \textit{Spitzer}) data and the $1.3\,{\rm mm}$ detection leaves
room for considering \isooph\ as a disk-bearing, envelope-free object, i.e., Class\,II source, 
as well. Our addition of \textit{Herschel} submm photometry has been fundamental to
removing this ambiguity. This makes the protostellar classification of ISO-Oph85 
reliable, compared to other published X-ray detected objects with presumed Class\,I status,
which have been classified using \textit{Spitzer} data. 
As a note of caution, we point out that there is a small offset between the \textit{HST}/NICMOS position of \isooph and the position of the emission peak in {\em Herschel}, 
even if it is  clearly within the {\em Herschel} positional errors. Even if we consider this scenario as quite unlikely, we cannot exclude that the emission observed in
{\em Herschel} might be associated with a starless core and not with \isooph. In this case, the millimeter identification of ISO-Oph85 by \cite{Motte98.0} would   also likely be erroneous.

Another important aspect is that the X-ray emission from ISO-Oph85 was detected
only as a result of our systematic time-resolved search for transient emission.
\cite{Pillitteri10.1} do not detect it in the same data set, based on time-averaged analysis,
which highlights the discovery space inherent in time-resolved source detection.

We calculate rough estimates for
$kT$ and $N_H$ from the analysis of energy quantiles: $kT = 1.15_{-0.65}^{+2.35}\,{\rm keV}$ e $N_H = 1.0_{-0.5}^{+1.2}\cdot 10^{23}\,{\rm cm^{-2}}$. 
At its lower end, the range of $N_H$ comprises  the value we
derive from the optical extinction $A_V =45^{+1}_{-3}$ that results from the SED fit. However, the
gas-to-dust extinction law in star-forming regions is notoriously uncertain (\citealt{2003A&A...408..581V}).
Using the above-mentioned spectral properties, 
we derive an X-ray luminosity of  $\log {L_X} = 31.1_{-1.2}^{+2.0}{\rm \,erg/s  }$ for the flare of ISO-Oph85. 
The upper limit representing the quiescent stage of ISO-Oph\,85 is
$\log{L_{\rm X}} <29.5, $  as derived using the $N_H$ and $kT$ value measured in the quantile diagram.
The spectral properties of the X-ray emitting plasma of ISO-Oph\,85 are poorly 
constrained owing to low photon statistics.
By making a comparison of the obtained temperature and absorption with  other YSOs in DROXO, as detected by \cite{Pillitteri10.1}, we
see that the flare emission from \isooph is located in the medium-temperature and high-absorption tail of the distribution.
In particular, its position is closer to the clustering region of the Class\,I and flat spectrum sources
than to the bulk of the Class\,II and Class\,III sources. 
We also note that Class\,I sources cluster near the boundary of our grid, at very high temperatures
 and high absorptions (see Fig. \ref{fig:sed}), and the same is true for `flat' sources.
Class\,II and especially Class\,III sources instead tend to cluster in the lower half of the grid, which  corresponds with lower
hydrogen column densities, and with a wide range of temperature. Comparing these results with the distribution of YSOs that were observed inside the Taurus Molecular Cloud
by \cite{2007A&A...468..405S}, we note that the barycenter of their distribution is located at
a lower value of hydrogen column density with respect to the barycenter of the distribution in the $\rho$ Ophiuchu cloud complex. This indicates a larger gas extinction in the $\rho$ 
Ophiuchi cloud, which is consistent with the larger dust extinction observed in the IR.

According to \cite{1997Natur.387...56G, 2005A&A...429..963O,Imanishi01.1}, and \cite{2003ApJ...582..398F}, the X-ray activity of a YSO 
roughly tends to increase as it gets older, moving from an earlier  YSO class to a later one, meaning that 
Class\,0 and I protostars have lower X-ray activity than  T Tauri stars (Class\,II and Class\,III). 
Among Class I protostars, X-ray emission has only been detected for a few objects, while 
the X-ray emission of Class\,0 objects  has yet to be confirmed.
The fraction of  X-ray detection  of Class\,I protostars is generally smaller
than for Class\,II and III. It is not clear if this is due to the high absorption,
as DROXO X-ray luminosity functions suggest, or because of intrinsically lower $L_X$. A third possibility for the low X-ray detection
rate of Class\,I objects, suggested by our result
for ISO-Oph85, is strong variability paired with low quiescent X-ray flux.

By adopting the submm peak as representative of the envelope of ISO-Oph85, the flare observed on \isooph confirms that  protostars can exhibit strong
X-ray activity. This behavior has  already been noticed e.g., by \cite{Imanishi01.1}. However, most X-ray variability studies of
protostars suffer from poor sensitivity
and provide access only to coarse variability measures such as Kolmogorov-Smirnov (KS) statistics 
and poor time-sampling (\citealt{Getman07.1}; \citealt{2014ApJS..213....4P}). 
The detection of a flare with $\sim2$ decades amplitude is, therefore, notable. 
Since we observed only one flare in $209.1\,{\rm ks}$ of observation, we can state that X-ray flares of this amplitude or larger only occur in \isooph   every $2.5\,{\rm d}$, roughly.
This result is consistent with the flaring frequency observed in the Taurus Molecular Cloud (TMC) and in the Orion Nebula Cloud (ONC) by \cite{2007A&A...468..463S}.
It is commonly thought that flares in embedded YSOs  originate from coronal magnetic reconnection that is due to a stellar
dynamo or to a  magnetic reconnection between the field lines of the YSO itself, of the YSO and the disk, or of the disk itself.
Since protostars are probably too young  to have internal dynamo-like dynamics that are capable of producing intense magnetic
reconnection from buoyant flux tubes \citep{2001EP&S...53..683H} -- 
as occurs in Sun-like stars -- X-ray flares on protostars might be an indicator that  the latter process dominates in X-ray emission. This could also mean 
that the stellar disk and possibly infalling envelopes have
strong magnetic fields  linked to that of the central forming protostar.

A systematic study of the X-ray flaring rate in $\rho$ Ophiuchi will be carried out in a separate paper, and will allow us to compare the flaring frequency in the $\rho$ Ophiuchi 
cloud complex with the results for TMC and ONC.  
A significant 
step forward in constraining the SED and understanding the YSO status could be made, for example, by using observations from the ALMA (\textit{Atacama Large Millimeter/submillimeter Array}, \citealt{2009IEEEP..97.1463W}).\\




\begin{acknowledgements}
We wish to thank Dr. Andrea Giuliani for useful discussions.
The research that led to these results has received funding
from the European Union Seventh Framework Programme
(FP7-SPACE-2013-1), under grant agreement n. 607452,
``Exploring the X-ray Transient and variable Sky -- EXTraS''.
\end{acknowledgements}

\bibliography{crbr51}
\bibliographystyle{aa}

\end{document}